# Deep Learning CNN for Pneumonia Detection: Advancing Digital Health in Society 5.0

**Hadi Almohab[1]***

[1]Faculty of Engineering, Computer, and Design, Nusa Putra University, Jalan Raya Cibolang, Sukabumi 43153, Jawa Barat, Indonesia

*Corresponding Author: hadi.almohab@nusaputra.ac.id

**Abstract:** Pneumonia merupakan masalah kesehatan global yang serius dan menyumbang tingkat morbiditas serta mortalitas yang tinggi, terutama di wilayah dengan keterbatasan alat diagnostik dan sumber daya kesehatan. Penelitian ini bertujuan mengembangkan model Convolutional Neural Network (CNN) berbasis deep learning untuk mendeteksi pneumonia secara otomatis menggunakan citra X-ray dada. Metode yang digunakan meliputi pelatihan model pada dataset berlabel dengan serangkaian teknik pra-pemrosesan, seperti normalisasi, augmentasi data, dan peningkatan kualitas citra untuk memperbaiki ketahanan dan kemampuan generalisasi model. Hasil pengujian menunjukkan bahwa model yang dioptimalkan mencapai akurasi uji 91,67%, dengan nilai ROC-AUC 0,96 dan PR-AUC 0,95, yang menandakan performa kuat dalam membedakan pneumonia dari citra normal. Kesimpulannya, model CNN ini memiliki potensi signifikan sebagai alat bantu diagnostik yang cepat, konsisten, dan andal, serta mendukung visi Society 5.0 dalam integrasi kecerdasan buatan untuk meningkatkan layanan kesehatan dan kesejahteraan masyarakat.

**Keywords:** CNN, Deep Learning, Kesehatan Digital, Pneumonia, Society 5.0, X-ray Dada

## PENDAHULUAN

Pneumonia merupakan salah satu penyakit pernapasan paling serius di dunia, menyebabkan jutaan kasus dan kematian setiap tahun, terutama pada anak-anak di bawah lima tahun, lansia, dan pasien dengan sistem kekebalan tubuh yang lemah(World Health Organization, 2019; UNICEF, 2018). Penyakit ini menyebabkan peradangan pada alveoli, yang mengakibatkan penumpukan cairan sehingga menghambat pertukaran gas dan dapat menimbulkan komplikasi pernapasan yang serius (Ruuskanen et al., 2011; Prina et al., 2015). Di negara berkembang, beban penyakit ini bahkan lebih tinggi karena kondisi hidup yang buruk, terbatasnya sumber daya kesehatan, dan keterlambatan penanganan medis, yang berkontribusi pada tingginya angka kematian (Rudan et al., 2008; Almohab & Al-Othmany, 2024). Sebagai salah satu penyebab kematian global utama, pneumonia tetap menjadi tantangan besar bagi kesehatan masyarakat (Rello & Diaz, 2003).

Diagnosis dini dan akurat sangat penting untuk mengurangi komplikasi dan mortalitas akibat pneumonia. Pencitraan X-ray dada (CXR) tetap menjadi metode diagnostik yang paling mudah diakses dan biaya-efektif(Neuman et al., 2012; Hendee & Ritenour, 2003). Namun, interpretasinya sangat bergantung pada radiolog, sehingga dapat menimbulkan variabilitas hasil dan kesalahan diagnosis (Kasban et al., 2015; Almohab et al., 2024). Masalah ini menjadi lebih kritis di wilayah dengan sumber daya terbatas, di mana kekurangan radiolog terampil menyebabkan keterlambatan atau ketidakakuratan diagnosis (Williams et al., 2013). Dengan kemajuan pesat kecerdasan buatan (AI), terutama machine learning (ML) dan deep learning (DL), muncul peluang baru untuk mendukung deteksi pneumonia menggunakan pencitraan medis (Witarsyah et al., 2025;Dixit, 2018). Convolutional Neural Networks (CNN) dikenal luas karena efektivitasnya dalam mengekstraksi dan mempelajari fitur dari citra mentah, serta mampu mencapai kinerja tinggi pada tugas klasifikasi (Lal et al., 2021;Pramanik et al., 2017). Transfer learning lebih lanjut meningkatkan performa CNN ketika bekerja dengan data berlabel terbatas (Hendee et al., 2010;Yu et al., 2018), sementara strategi ensemble memperkuat ketahanan model dengan mengurangi bias dari model individu (Dalhoumi et al., 2015).

Meski demikian, sebagian besar metode yang ada bergantung pada teknik pra-pemrosesan seperti Dynamic Histogram Equalization (DHE) atau Histogram Equalization (HE) untuk meningkatkan kualitas citra. Teknik-teknik ini memang dapat memperbaiki kontras, namun juga berpotensi menimbulkan artefak atau noise yang menurunkan keandalan hasil klinis. Kesenjangan ini menunjukkan perlunya metode yang mempertahankan akurasi tanpa memerlukan pra-pemrosesan yang kompleks. Untuk mengatasi masalah tersebut, penelitian ini mengembangkan sistem diagnosis berbantuan komputer (CAD) berbasis CNN untuk mengklasifikasikan citra CXR menjadi kategori normal dan pneumonia tanpa menerapkan langkah pra-pemrosesan lanjutan. Kinerja model dievaluasi pada berbagai epoch pelatihan menggunakan metrik akurasi, presisi, recall, F1-score, dan kurva Receiver Operating Characteristic (ROC). Kebaruan penelitian ini terletak pada demonstrasi bahwa CNN dapat mencapai akurasi diagnostik tinggi tanpa bergantung secara berat pada teknik pra-pemrosesan. Inovasi ini membuat sistem ini sangat relevan untuk lingkungan kesehatan dengan sumber daya terbatas, mendukung deteksi pneumonia dini, dan berkontribusi pada peningkatan perawatan pasien sesuai visi Society 5.0.

## METODE

### Waktu dan Tempat Penelitian

Penelitian ini dilaksanakan pada tahun 2025 dengan menggunakan *Chest X-ray Pneumonia Dataset* yang tersedia secara publik di Kaggle. Dataset ini dikumpulkan dari anak-anak berusia 1–5 tahun di Guangzhou Women and Children's Medical Center, Tiongkok.

### Populasi dan Sampel

Data terdiri dari total 5.863 citra X-ray dada anak-anak dengan label Pneumonia atau Normal. Dataset dibagi menjadi 5.216 citra untuk pelatihan, 16 citra untuk validasi, dan 624 citra untuk pengujian. Kualitas, keberagaman, dan penggunaan luas dataset ini menjadikannya standar benchmark dalam penelitian deteksi pneumonia. Sumber Dataset: Chest X-ray Pneumonia Dataset on Kaggle

### Prosedur Penelitian

Penelitian ini menggunakan Convolutional Neural Network (CNN) berbasis deep learning dan mengikuti alur kerja terstruktur yang meliputi persiapan dataset, pra-pemrosesan data, perancangan model, pelatihan, dan evaluasi. Semua citra dikonversi ke skala abu-abu, diubah ukurannya menjadi 128 × 128 piksel, dan dinormalisasi ke rentang [0, 1]. Untuk meningkatkan generalisasi dan mengurangi overfitting, dilakukan data augmentation pada dataset pelatihan, termasuk rotasi acak (±30°), zoom hingga 20%, flipping horizontal, serta pergeseran lebar/tinggi kecil menggunakan Keras ImageDataGenerator. Model CNN terdiri dari dua blok konvolusi dengan 64 dan 128 filter masing-masing diikuti max pooling 2×2, flatten layer, dan dense layer 128 neuron dengan aktivasi ReLU serta dropout 0,5. Layer keluaran menggunakan satu neuron dengan aktivasi sigmoid untuk klasifikasi biner Pneumonia vs. Normal. Model dikompilasi menggunakan optimizer Adam (learning rate 0,001) dan binary cross-entropy loss dengan akurasi sebagai metrik evaluasi. Callback ReduceLROnPlateau digunakan untuk menurunkan learning rate bila validation loss tidak meningkat selama tiga epoch berturut-turut (minimum LR = $1\times10^{-5}$). Pelatihan dilakukan pada dataset teraugmentasi selama 10, 20, dan 50 epoch dengan batch size 32.

### Teknik Analisis Data

Model yang telah dilatih dievaluasi pada *test set* independen menggunakan metrik akurasi, presisi, *recall*, dan F1-score. Kurva Receiver Operating Characteristic (ROC) dengan AUC dan kurva Precision–Recall (PR) digunakan untuk menilai kemampuan diskriminatif model. *Confusion matrix* dihasilkan untuk memberikan wawasan mengenai kesalahan klasifikasi antara kasus Pneumonia dan Normal.

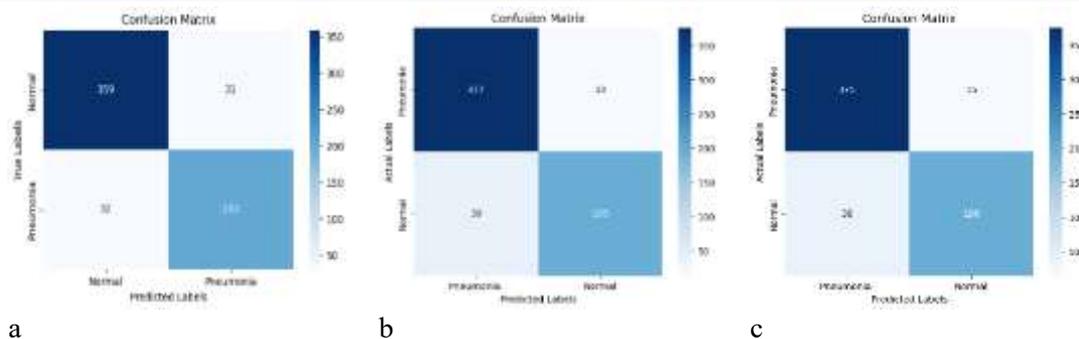

**Gambar 1.** Matriks kebingungan model CNN untuk berbagai durasi pelatihan: (a) 10 epoch, (b) 20 epoch, dan (c) 50 epoch

## HASIL DAN PEMBAHASAN

Bagian ini menyajikan hasil eksperimen serta memberikan pembahasan rinci terkait kinerja model yang dikembangkan.

### Hasil

Model CNN yang diusulkan dievaluasi pada 624 citra X-ray dada dengan tiga durasi pelatihan berbeda: 10, 20, dan 50 epoch. Tabel 1 merangkum akurasi uji, presisi, recall, dan F1-score untuk kelas Normal dan Pneumonia. Akurasi uji tertinggi, yaitu 91,67%, diperoleh pada pelatihan 20 epoch. Model menunjukkan kinerja yang konsisten pada kedua kelas, dengan presisi dan recall yang sedikit lebih tinggi pada kelas Normal, yang mengindikasikan performa klasifikasi yang kuat dan stabil. Matriks kebingungan (confusion matrix) memberikan wawasan tambahan terkait kinerja klasifikasi pada setiap epoch. Pada 10 epoch, model masih salah mengklasifikasikan lebih banyak sampel (31 citra Normal Pneumonia dan 32 citra Pneumonia Normal). Kinerja meningkat secara signifikan pada **20** epoch, di mana hanya 13 citra Normal dan 39 citra Pneumonia yang salah klasifikasi. Pada 50 epoch, hasilnya relatif serupa, dengan 15 citra Normal dan 38 citra Pneumonia yang salah klasifikasi.Temuan ini menunjukkan bahwa model mencapai keseimbangan terbaik antara sensitivitas dan spesifisitas pada 20 epoch. Kurva akurasi dan loss pada data pelatihan serta validasi semakin memperjelas dinamika proses pembelajaran. Seperti ditunjukkan pada Gambar 2, akurasi meningkat secara bertahap selama pelatihan, sementara nilai loss menurun secara konsisten, yang mengindikasikan konvergensi model yang efektif pada setiap epoch.

**Table 1.** Ringkasan Kinerja Model CNN pada Data Uji

| Epoch | Akurasi Uji (%) | Kelas | Presisi | Recall | F1-Score |
|---|---|---|---|---|---|
| 10 | 89,90 | Normal | 0,92 | 0,92 | 0,92 |
|  |  | Pneumonia | 0,87 | 0,86 | 0,87 |
| 20 | 91,67 | Normal | 0,91 | 0,97 | 0,94 |
|  |  | Pneumonia | 0,94 | 0,83 | 0,88 |
| 50 | 91,51 | Normal | 0,91 | 0,96 | 0,93 |
|  |  | Pneumonia | 0,93 | 0,84 | 0,88 |

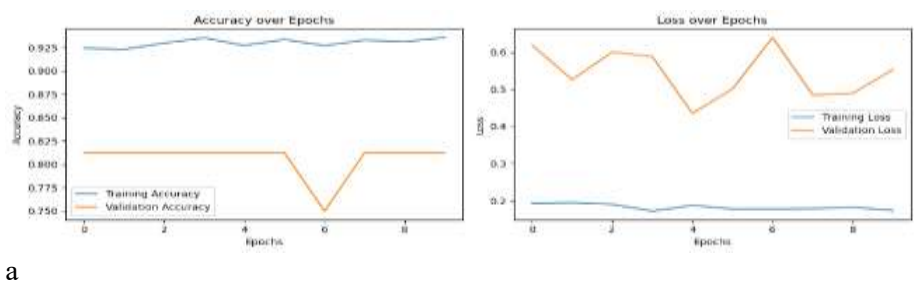

a

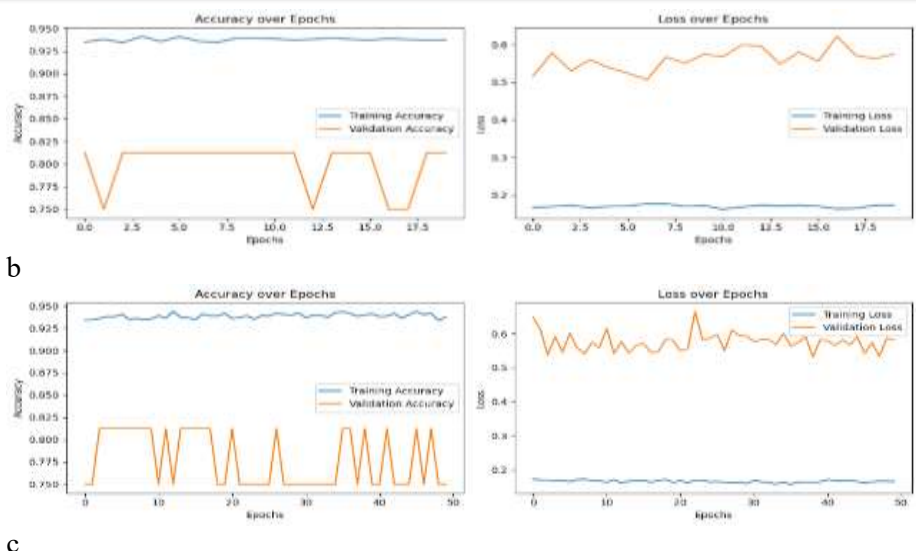

b

c

**Gambar 2.** Kurva akurasi dan loss pelatihan serta validasi dari model CNN pada durasi pelatihan berbeda: (a) 10 epoch, (b) 20 epoch, dan (c) 50 epoch.

Kinerja diskriminatif model semakin tervalidasi melalui kurva Precision-Recall (PR) dan Receiver Operating Characteristic (ROC). Seperti ditunjukkan pada Gambar 3, kurva PR mencapai AUC sebesar 0,95 (a), sedangkan kurva ROC mencapai AUC sebesar 0,96 (b). Hasil ini menegaskan kemampuan model yang kuat dalam membedakan kasus Normal dan Pneumonia.

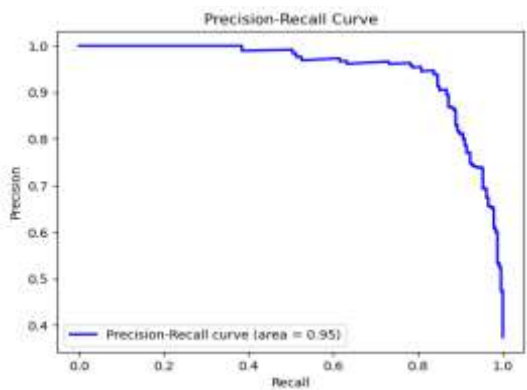
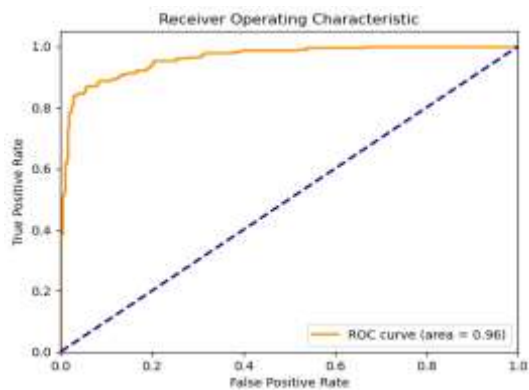

a  b

**Gambar 3.** Kinerja model CNN pada dataset uji: (a) kurva Precision–Recall (AUC = 0,95) dan (b) kurva ROC (AUC = 0,96).

Untuk memberikan konteks terhadap hasil penelitian, Tabel 2 membandingkan model CNN yang diusulkan dengan beberapa penelitian sebelumnya. Model yang diusulkan menunjukkan kinerja yang kompetitif dalam hal akurasi, presisi, recall, dan AUC, menegaskan efektivitasnya dalam deteksi pneumonia.

**Tabel 2.** Perbandingan kinerja model CNN dengan penelitian sebelumnya

| Model [Ref.] | Akurasi Uji (%) | Presisi | Recall | AUC |
| --- | --- | --- | --- | --- |
| Model (Liang & Zheng, 2020) | 90.50 | 0.8910 | 0.9670 | – |
| Model1 (Jain et al., 2020) | 85.26 | 0.7500 | 0.9400 | – |
| DenseNet121 (Original) (Zhang et al., 2021) | 91.37 | 0.7694 | 0.9715 | 0.9845 |
| DenseNet121 (Enhanced) (Zhang et al., 2021) | 93.42 | 0.8041 | 1.0000 | 0.9957 |
| Proposed CNN Model (20 epochs) | 91.67 | 0.91 | 0.90 | 0.96 |

**Pembahasan**

Hasil penelitian menunjukkan bahwa model CNN yang diusulkan mampu mengklasifikasikan citra X-ray dada menjadi kategori Normal dan Pneumonia secara efektif. Peningkatan jumlah epoch pelatihan dari 10 menjadi 20 meningkatkan kinerja model, sedangkan pelatihan hingga 50 epoch hanya memberikan peningkatan yang marginal, menunjukkan bahwa konvergensi stabil tercapai sekitar 20 epoch. Analisis matriks kebingungan menunjukkan bahwa sebagian besar kesalahan klasifikasi terjadi pada kasus Pneumonia yang diprediksi sebagai Normal, sesuai dengan recall yang sedikit lebih rendah pada kelas Pneumonia. Pola ini umum terjadi pada dataset dengan ketidakseimbangan kelas yang moderat. Meskipun demikian, matriks kebingungan menegaskan bahwa secara keseluruhan kesalahan klasifikasi minimal, dan model mampu menggeneralisasi dengan baik pada data uji yang belum pernah dilihat sebelumnya. Kurva PR dan ROC semakin mendukung ketahanan model, dengan nilai AUC tinggi yang menunjukkan kemampuan diskriminatif yang sangat baik. Secara keseluruhan, temuan ini menunjukkan bahwa pra-pemrosesan yang tepat, normalisasi, dan data augmentation, dikombinasikan dengan arsitektur CNN, dapat mendeteksi pneumonia dari citra X-ray dada secara andal. Selain itu, analisis perbandingan pada Tabel 2 menekankan bahwa model CNN yang diusulkan menunjukkan kinerja yang kompetitif dibandingkan dengan pendekatan mutakhir lainnya.

**KESIMPULAN**

Penelitian ini berhasil mengembangkan dan mengevaluasi model Convolutional Neural Network (CNN) berbasis deep learning untuk deteksi pneumonia secara otomatis menggunakan citra X-ray dada, dengan kinerja diagnostik yang kuat ditunjukkan oleh akurasi uji 91,67% pada 20 epoch pelatihan serta nilai ROC-AUC 0,96 dan PR-AUC 0,95, yang mencerminkan kemampuan diskriminatif yang tinggi dan tingkat salah klasifikasi yang minimal. Temuan ini membuktikan bahwa arsitektur CNN yang dirancang secara tepat, dikombinasikan dengan normalisasi dan augmentasi data, mampu mencapai performa kompetitif tanpa memerlukan teknik pra-pemrosesan kompleks yang berpotensi menimbulkan artefak, sehingga model tetap relevan untuk diterapkan pada lingkungan kesehatan dengan keterbatasan sumber daya karena dapat memberikan dukungan diagnostik yang cepat, konsisten, dan andal. Selaras dengan visi Society 5.0, hasil penelitian ini mempertegas potensi integrasi kecerdasan buatan untuk meningkatkan aksesibilitas dan efisiensi layanan kesehatan, sementara arah penelitian selanjutnya meliputi perluasan dataset dengan variasi demografi dan etiologi pneumonia serta pengembangan implementasi klinis melalui integrasi model dengan sistem informasi rumah sakit guna memastikan dampak nyata di lapangan.

**UCAPAN TERIMA KASIH**

**REFERENSI**